\documentclass[11pt]{article}
\usepackage{moriond,epsfig}

\def\be{\begin{equation}}
\def\ee{\end{equation}}
\def\bea{\begin{eqnarray}}
\def\eea{\end{eqnarray}}

\def\ga{\mathrel{\hbox{\rlap{\hbox{\lower4pt\hbox{$\sim$}}}\hbox{$>$}}}}
\def\la{\mathrel{\hbox{\rlap{\hbox{\lower4pt\hbox{$\sim$}}}\hbox{$<$}}}}
\def\arcsec{\hbox{$^{\prime\prime}$}}

\begin{document}
\noindent To appear in {\it Proceedings of XXI Moriond Conference:
Galaxy Clusters and the High Redshift Universe Observed in X-rays},
edited by D. Neumann, F. Durret, \& J. Tran Thanh Van, in press
\vspace*{3.0cm}
\title{MERGER SHOCKS AND NONTHERMAL PROCESSES IN CLUSTERS OF GALAXIES}

\author{CRAIG L. SARAZIN}

\address{Department of Astronomy, University of Virginia, P. O. Box 3818, \\
 Charlottesville, VA 22903-0818, U.S.A.}

\maketitle\abstracts{Clusters of galaxies generally form by the
gravitational merger of smaller clusters and groups.
Major cluster mergers are the most energetic events in the Universe
since the Big Bang.
Mergers drive shocks into the intracluster gas, and
these shocks heat the intracluster gas, and should also accelerate
nonthermal relativistic particles.
The thermal effects of merger shocks will be briefly discussed.
Mergers can increase the temperature and X-ray luminosities of
clusters, and this may affect statistical conclusions about
cosmological parameters.
As a result of particle acceleration in shocks,
clusters of galaxies should contain very large populations of
relativistic electrons and ions.
Electrons with Lorentz factors $\gamma \sim 300$
(energies $E = \gamma m_e c^2 \sim 150$ MeV) are expected to
be particularly common.
Observations and models for the radio, extreme
ultraviolet, hard X-ray, and gamma-ray emission from nonthermal
particles accelerated in these shocks will also be described.
The predicted gamma-ray fluxes of clusters should make them easily
observable with GLAST.
Chandra X-ray observations of the interaction between the radio lobes and
cooling flow gas in Abell~2052 are also discussed briefly.}

\section{Introduction}\label{sec:intro}

\begin{figure}[t]
\vskip3.65truein
\includegraphics{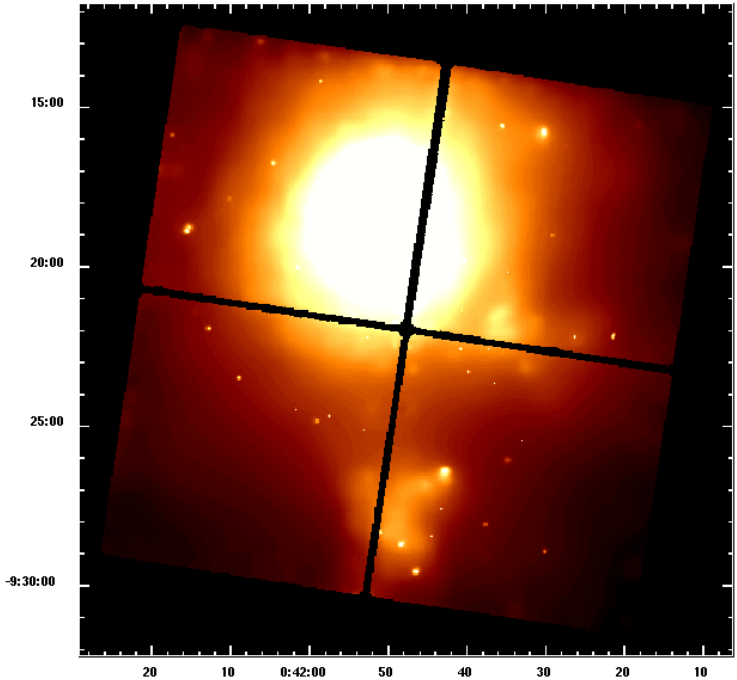}
\caption[]{The Chandra ACIS-I image of the merger cluster Abell~85.\cite{KSR}
The false color scale burns out the central cooling flow region to show the
outer parts of the cluster.
Two subclusters to the south and southwest are merging with the main
cluster.
The southwestern subcluster has a radio relic, and the southern subcluster
also has a diffuse radio source which may be a relic.\cite{bagchi}
The sharp feature at the northwest of the southern subcluster is a
``cold front''.\cite{Vea}\hfill
\label{fig:a85}}
\end{figure}

Major cluster mergers are the most energetic events in the Universe
since the Big Bang.
Cluster mergers are the mechanism by which clusters are assembled.
In these mergers, the subclusters collide at velocities of
$\sim$2000 km/s,
releasing gravitational binding energies of as much as $\ga$$10^{64}$
ergs.
Figure~\ref{fig:a85} shows the Chandra image of the merging cluster
Abell~85, which has two subclusters merging with the main cluster.\cite{KSR}
Spectral measurements show that the southern subcluster is merging
at a relative velocity of about 2800 km/s.

The relative motions in mergers are moderately supersonic,
and shocks are driven into the intracluster medium.
In major mergers, these hydrodynamical shocks dissipate energies of
$\sim 3 \times 10^{63}$ ergs; such shocks are the major heating
source for the X-ray emitting intracluster medium.
The shock velocities in merger shocks are similar to those in
supernova remnants in our Galaxy, and we expect them to produce similar
effects.
Mergers shocks should heat and compress the X-ray emitting intracluster
gas, and increase its entropy.
We also expect that particle acceleration by these shocks will produce
nonthermal electrons and ions, and these can produce synchrotron
radio, inverse Compton (IC) EUV and hard X-ray, and gamma-ray emission.

\begin{figure}[t]
\vskip3.8truein
\includegraphics{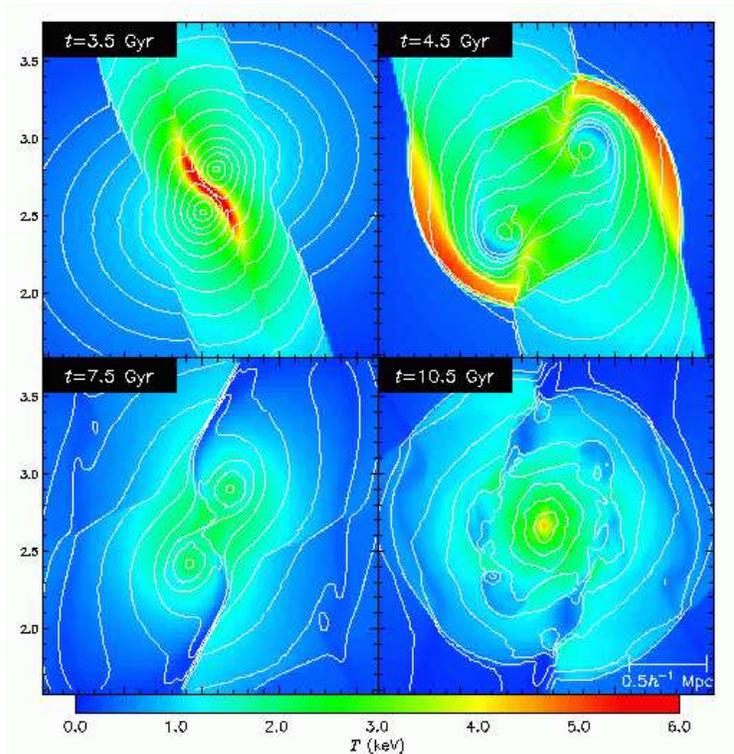}
\caption[]{The results of a hydro simulation of a symmetric, off-center
merger by Ricker and Sarazin.\protect\cite{RS}
The colors show the temperature, while the contours are the X-ray surface
brightness.
Initially, the shocked region is located between the two subcluster
centers.
Later, the main merger shocks propogate to the outer parts of the cluster,
and other weaker shocks also occur.
By the end of the simulation, the cluster is beginning to return to
equilibrium.\hfill
\label{fig:merger_color}}
\end{figure}

Hydrodynamical simulations of cluster formation and evolution have shown
the importance of merger shocks.\cite{SM,RSB}
The evolution of the structure of merger shocks is illustrated in
Figure~\ref{fig:merger_color},\cite{RS}
which shows an off-center merger between two symmetric subclusters.
At early stages in the merger (the first panel and earlier), the shocked
region is located between the two subcluster centers and is bounded
on either side by two shocks.
At this time, the subcluster centers, which may contain cooling cores
and central radio sources, are not affected.
Later, these shocks sweep over the subcluster centers (between the first and
second panels);
the survival of central cool cores depends on how sharply peaked the
potentials of the subclusters are.\cite{MSV,RS}
The main merger shocks pass into the outer parts of the merging system
(panel 2),
and secondary shocks may appear in the inner regions (panel 3).
Eventually, the cluster begins to return to equilibrium (panel 4).

\section{Thermal Effects of Merger Shocks}\label{sec:thermal}

Merger shocks heat and compress the intracluster gas, and these effects
can be used to determine the geometry and kinematics of the merger.\cite{MSV}
Recently, Chandra images have detected a number of merger shock
regions in clusters.\cite{Mark,MVMV}
Merger shocks can also mix the intracluster medium.

\subsection{Mergers and Cooling Flows} \label{sec:thermal_cf}

\begin{figure}[t]
\includegraphics{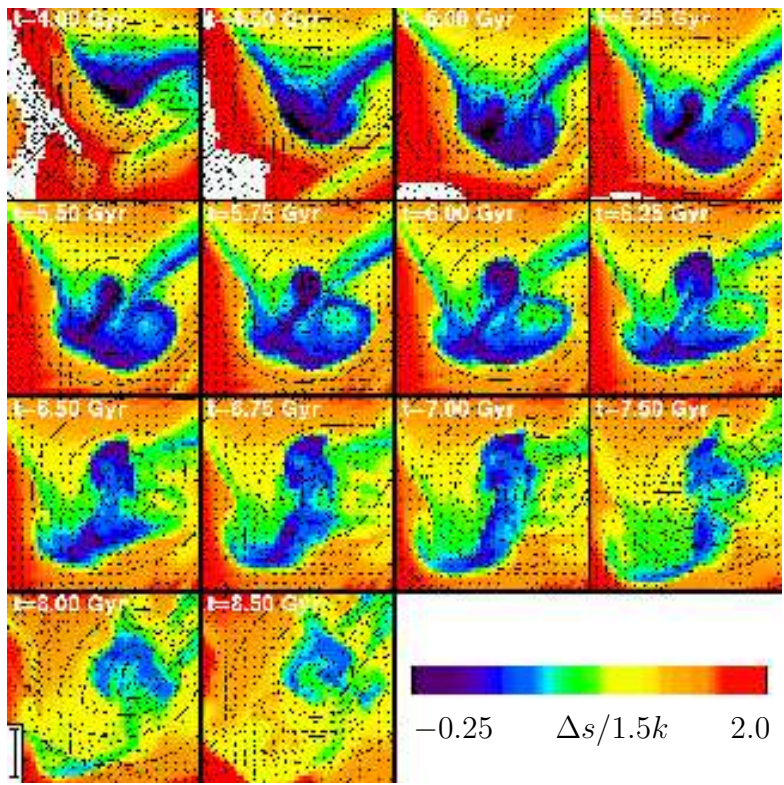}
\centerline{\null}
\vskip3.8truein
\noindent\hskip3.3truein
{\large $-0.25 ~~~~~~ \Delta s / 1.5k ~~~~~~ 2.0$}
\vskip0.2truein
\caption[]{A merger-driven convective plume mixing the low entropy gas
at the core of a cluster with higher entropy surrounding gas.\cite{RS}
The plots show the specific entropy (entropy per particle) in units of
$(3/2)k$ in the inner $\sim$600 $h^{-1}$ kpc square of a merging cluster
simulation at a series of time steps covering about 4.5 Gyr.
Blue colors are low entropy gas, while  red and white colors are high
entropy.\hfill
\label{fig:mixing}}
\end{figure}

Merger shocks also appear to disrupt cooling flows.
There is a strong statistical anticorrelation between cooling flows
and/or cooling rates,
and irregular structures in clusters as derived by statistical analysis
of their X-ray images.\cite{BT}
It is unlikely that this is due to shock heating of the cooling flow
gas.\cite{FD,gomez,RS,S2001}
Numerical simulations and physical arguments suggest that the disruption may
be due to the ram pressure of the gas from the other
subcluster.\cite{FD,gomez,MSV,RS,S2001}
The cooling flow gas is displaced from the potential minimum, and
eventually mixes with hotter, more diffuse gas.
Figure~\ref{fig:mixing} shows the results of a numerical simulation of
an offset merger between subclusters with a mass ratio of about 1:3.\cite{RS}
The color scale shows the specific entropy of the gas in the center of one
of the subclusters.
Initially, the ram pressure of the merging subcluster gas displaces the
lower entropy gas from the center of the potential, causing it to become
convectively unstable.
The resulting convective plumes produce large-scale
turbulent motions with eddy sizes up to several 100~kpc.
This mixing eventually eliminates the lower entropy gas at the cluster
core.

Assuming that ram pressure is the main mechanism for disrupting cooling
flows in mergers, one expects that the merger will remove the cooling flow
gas at radii which satisfy
\begin{equation} \label{eq:ram}
\rho_{\rm sc} v_{\rm rel}^2 \ga P_{\rm CF} ( r ) \, ,
\end{equation}
where $P_{\rm CF} ( r )$ is the pressure profile in the cooling flow,
$\rho_{\rm sc}$ is the density of the merging subcluster gas at
the location of the cooling flow, and $v_{\rm rel}$ is the relative
velocity of the merging subcluster gas and the cooling flow.
G\`omez et al.\ (2001) find that this relation provides a reasonable
approximation to the disruption in their simulations.
The pressure profile in the cooling flow gas prior to the merger is
determined by the condition of hydrostatic equilibrium.
If the cluster gravitational potential has a wide core within which
the potential is nearly constant
(e.g., as in a King model),
then the cooling flow pressure will not increase rapidly into the center.
In this case, once the merger reaches the central regions of the cluster,
if the ram pressure is sufficient to remove the outer parts of the
cooling flow, it should be sufficient to remove nearly all of the cooling
flow.
On the other hand, if the cluster potential is sharply peaked
(as in a NFW profile\cite{NFW}),
the merger may remove the
outer parts of the cooling flow but not the innermost regions.
Thus, the survival and size of cool cores in merging clusters
can provide evidence on whether clusters have sharply peaked
potentials.\cite{MSV}

\subsection{Luminosity and Temperature Boosts in Mergers}
\label{sec:thermal_boost}

\begin{figure}[t]
\vskip3.6truein
\includegraphics{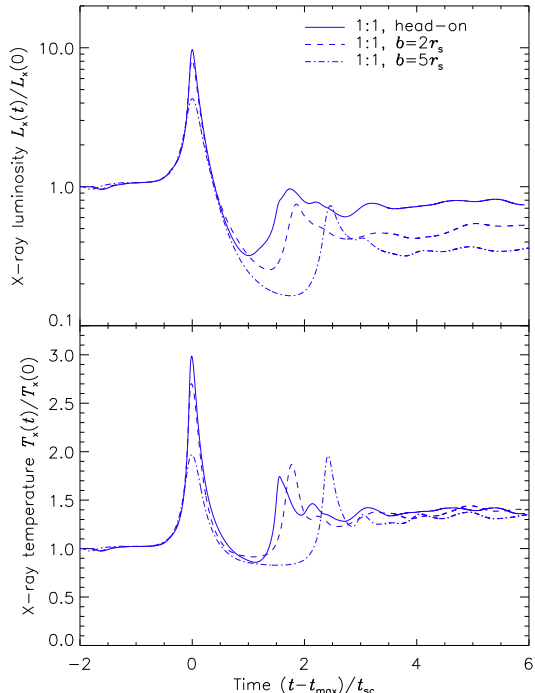}
\caption[]{The X-ray luminosity $L_X$ (2--10 keV) and average
emission-weighted X-ray temperature $T_X$ as functions of time $t$
in merger simulations for equal mass subclusters.\cite{RS}
The luminosity and temperature are for the total system (both subclusters)
and are scaled to their initial values.
Time is offset from the time of peak luminosity $t_{\rm max}$ and
scaled to the sound-crossing time $t_{\rm sc}$ for each subcluster.
The three curves are for different values of the impact parameter $b$ for
the collision.\hfill
\label{fig:boost}}
\end{figure}

The compression and heating associated with merger shocks boosts the
X-ray luminosity and temperature of the intracluster gas.
Figure~\ref{fig:boost} shows the variation in the 2--10~keV X-ray
luminosity $L_X$ and the average emission-weighted X-ray temperature
$T_X$ as functions of the time $t$ in simulations of the merger of equal mass
subclusters.\cite{RS}
The three curves are for different values for the impact parameter for
the collision;
the solid curve is for a head-on collision.
The values of $L_X$ and $T_X$ are normalized to the sum
of the two subclusters prior to the merger.
The merger can increase the X-ray luminosity of the sum of the two
subclusters by a factor of $\sim$10 ($\sim$20 times the initial individual
luminosities of the subclusters).
The X-ray temperature can be increased by a factor of $\sim$3.

The most massive, luminous, and hottest clusters are relatively rare.
For relaxed, virialized clusters, the Press-Schechter formalism predicts
that the abundance of clusters drops exponentially with mass for the
largest systems.\cite{LC,PS}
The X-ray luminosity and temperature boosts during rare major merger
events are large enough to strongly affect the statistics of the hottest
and most X-ray luminous clusters.
Thus, one might expect that the most X-ray luminous and hottest
clusters would mainly be mergers.\cite{RSR}
There are a number of examples which support this argument.
The cluster 1E~0657-56 $=$ RXJ0658-5557 is probably the hottest known
cluster, with a temperature of $kT \approx 16$ keV.\cite{tucker,Mark}
The Chandra image of this cluster reveals a complex merger
with a spectacular bow shock.\cite{Mark}
Similarly, Abell 2163 may be the second hottest known cluster;
its Chandra image shows two apparent shock regions.\cite{MV}
Abell~665, one of the most X-ray luminous clusters
at redshift $z \la 0.2$, is currently undergoing a merger,
and may be near the time of core collision\cite{gomez2}
when the $L_X$ and $T_X$ boosts are the largest.\cite{RS}
The Chandra image shows a dramatic merger shock.\cite{Mark,MV}
The very distant, luminous, and hot cluster MS1054-03
shows evidence of substructure indicative of an ongoing merger.\cite{NA}

There is a strong, steep correlation between the radio luminosity of
clusters with radio halos, and their X-ray temperature and/or
luminosity.\cite{LHBA,Fer2}
This relationship is steeper than might be expected from simple
scaling arguments in clusters.\cite{KS2}
Since radio halos are only observed in clusters undergoing
mergers\cite{Fer1,SB},
and may be powered by merger shock acceleration
(\S~\ref{sec:nonthermal_radio}),
the strong correlation with the X-ray temperature and luminosity may be
due, in part, to the merger boost in these quantities.
This might explain why radio observations of Sunyaev-Zel'dovich clusters,
which are generally selected to be brightest and hottest clusters,
have found the (otherwise relatively rare) radio halos in a number of
cases.\cite{MB,LHBA}

The abundance of massive clusters at moderately large redshifts has
been used as an important diagnostic for the matter density in the
Universe and $\Omega_M$.\cite{OB,VL,BF,DV}
Since it is difficult to determine the mass directly, the X-ray temperature
or luminosity is often used in its place, and the observations are usually
analyzed assuming that the clusters are relaxed and virialized.
In Press-Schechter theory,
the abundance of massive clusters declines rapidly with mass
and with redshift, and the rate of decline with redshift increase
rapidly with $\Omega_M$.
Thus, the presence of hot, X-ray luminous clusters at significant
redshifts (e.g., MS1054-03) provides an upper limit on $\Omega_M$
which excludes the Einstein-de Sitter Universe.\cite{BF,DV}
However, if most of these clusters are undergoing mergers,
their masses may be overestimated and their abundance underestimated
by these equilibrium arguments.
Thus, mergers may bias the inferred value of $\Omega_M$ to lower
values.\cite{RSR}

\section{Particle Acceleration and Nonthermal Emission} \label{sec:nonthermal}

\subsection{Shock Acceleration} \label{sec:nonthermal_acceler}

Radio observations of supernova remnants indicate that
shocks with $v \ga 10^3$ km/s convert at least a few
percent of the shock energy into the acceleration of relativistic
electrons.\cite{BE}
Even more energy may go into relativistic ions.
While the merger shocks in cluster have lower Mach numbers and compressions
than the blast waves in young supernova remnants, it is probable that
shock acceleration operates efficiently in clusters as well.
Given that all of the thermal energy of the intracluster gas in clusters
is due to shocks with such velocities, it seems likely that relativistic
electrons with a total energy of $\ga 10^{62}$ ergs are produced in
clusters, with perhaps even higher energies in ions.

\begin{figure}[t]
\vskip2.30in
\includegraphics{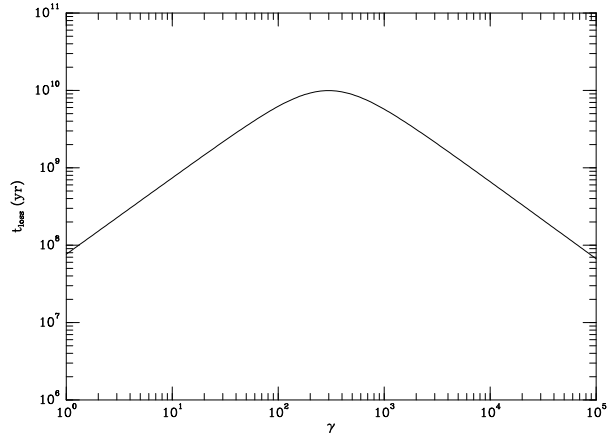}
\caption[]{The loss time scale for relativistic electrons in
a cluster of galaxies under typical intracluster conditions as
a function of their Lorentz factor $\gamma$.
The particles have energies of $E = \gamma m_e c^2$.\hfill
\label{fig:lifetimes}
}
\end{figure}

Clusters are also very good storage locations for cosmic rays.
Under reasonable assumptions for the diffusion coefficient, particles
with energies $\la$$10^6$ GeV have diffusion times which
are longer than the Hubble time.\cite{BBP,CB}
Figure~\ref{fig:lifetimes} gives the loss time scales for relativistic
electrons under typical intracluster conditions.
Although high energy electrons lose energy rapidly due to IC and
synchrotron emission, electrons with Lorentz factors of $\gamma \sim 300$
(energies $\sim 150$ MeV)
have long lifetimes which approach the Hubble time.\cite{S1,SL}
Thus, clusters of galaxies can retain low energy electrons
($\gamma \sim 300$) and nearly all cosmic ray ions for a significant
fraction of a Hubble time.

I have calculated models for the relativistic electrons in
clusters, assuming they are primary electrons accelerated in merger
shocks.\cite{S1,S2}
An alternative theory is that these particles are secondaries
produced by the decay of charged mesons generated in cosmic ray
ion collisions.\cite{CB}
Two recent cluster merger simulations have included particle acceleration
approximately.\cite{RBS,TN}
Their conclusions are very similar to mine based on simpler models.
The populations of cosmic ray electrons in clusters depends on their merger
histories.
Because low energy electrons have long lifetimes, one expects to find
a large population of them in most clusters (any cluster which has had
a significant merger since $z \sim 1$).
On the other hand, higher energy electrons ($E \ga 1$ GeV) have
short lifetimes (shorter than the time for a merger shock to cross
a cluster).
Thus, one only expects to find large numbers of higher energy electrons
in clusters which are having or have just had a merger.

Fig.~\ref{fig:espect}(a) shows the electron spectrum in a cluster
with a typical history.
Most of the electron energy is in electrons with $\gamma \sim 300$, which
have the longest lifetimes.
These electrons are produced by mergers over the entire history
of the cluster.
This cluster also has a small ongoing merger which produces
the high energy tail on the electron distribution.

Energetically, most of the emission from these electrons is due to IC,
and the resulting spectrum is shown in Fig.~\ref{fig:espect}(b).
For comparison, thermal bremsstrahlung with a typical rich cluster
temperature and luminosity is shown as a dashed curve.
Fig.~\ref{fig:espect}(b) shows that clusters should be strong
sources of extreme ultraviolet (EUV) radiation.
Since this emission is due to electrons with $\gamma \sim 300$ which
have very long lifetimes, EUV radiation should be a common feature
of clusters.\cite{SL}

In clusters with an ongoing merger, the higher energy electrons will
produce a hard X-ray tail via IC scattering of the Cosmic Microwave
Background (CMB);
the same electrons will produce diffuse radio synchrotron emission.

\subsection{EUV/Soft X-ray Emission} \label{sec:nonthermal_euv}

Excess EUV emission has apparently been detected with the EUVE satellite in
six clusters (Virgo, Coma, Abell 1795, Abell 2199, Abell 4038, \&
Abell 4059).\cite{Lea1,Lea2,BB,MLL,BBK2,Kea,Lea3,Lea4,BBK1,BLM,BBK3}
In fact, the EUVE satellite appears to have detected all of the clusters
it observed which are nearby, which have long integration times, and 
which lie in directions of low Galactic column where detection is
possible at these energies.
However, the EUV detections and claimed properties of the clusters
remain quite controversial.\cite{BB,AB,BBK2,BBK1}

The EUV observations suggest that rich clusters generally have
EUV luminosities of $\sim$$10^{44}$ ergs/s, and have spectra which
decline rapidly in going from the EUV to the X-ray band.
The EUV emission is very spatially extended.

\begin{figure}[t]
\includegraphics{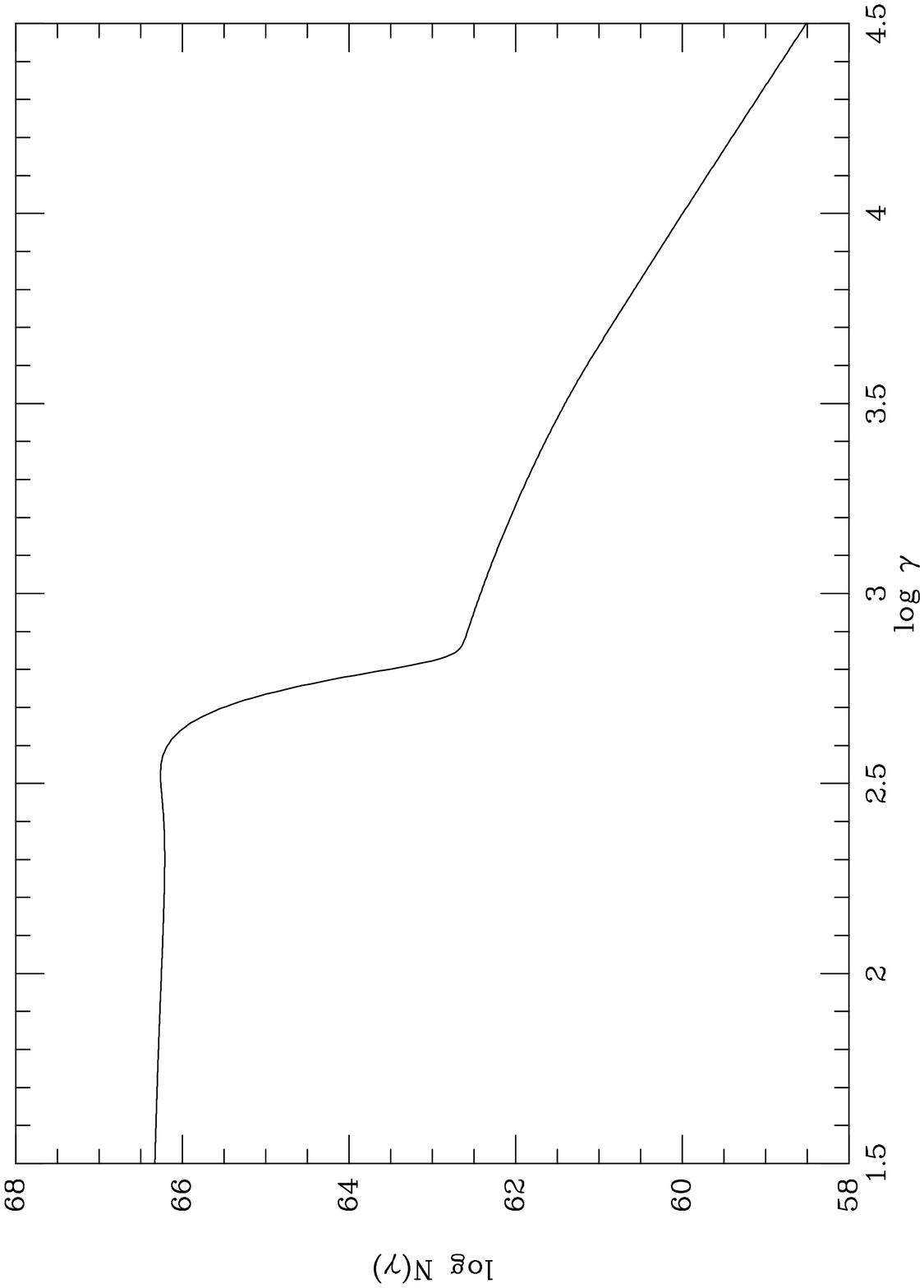}
\includegraphics{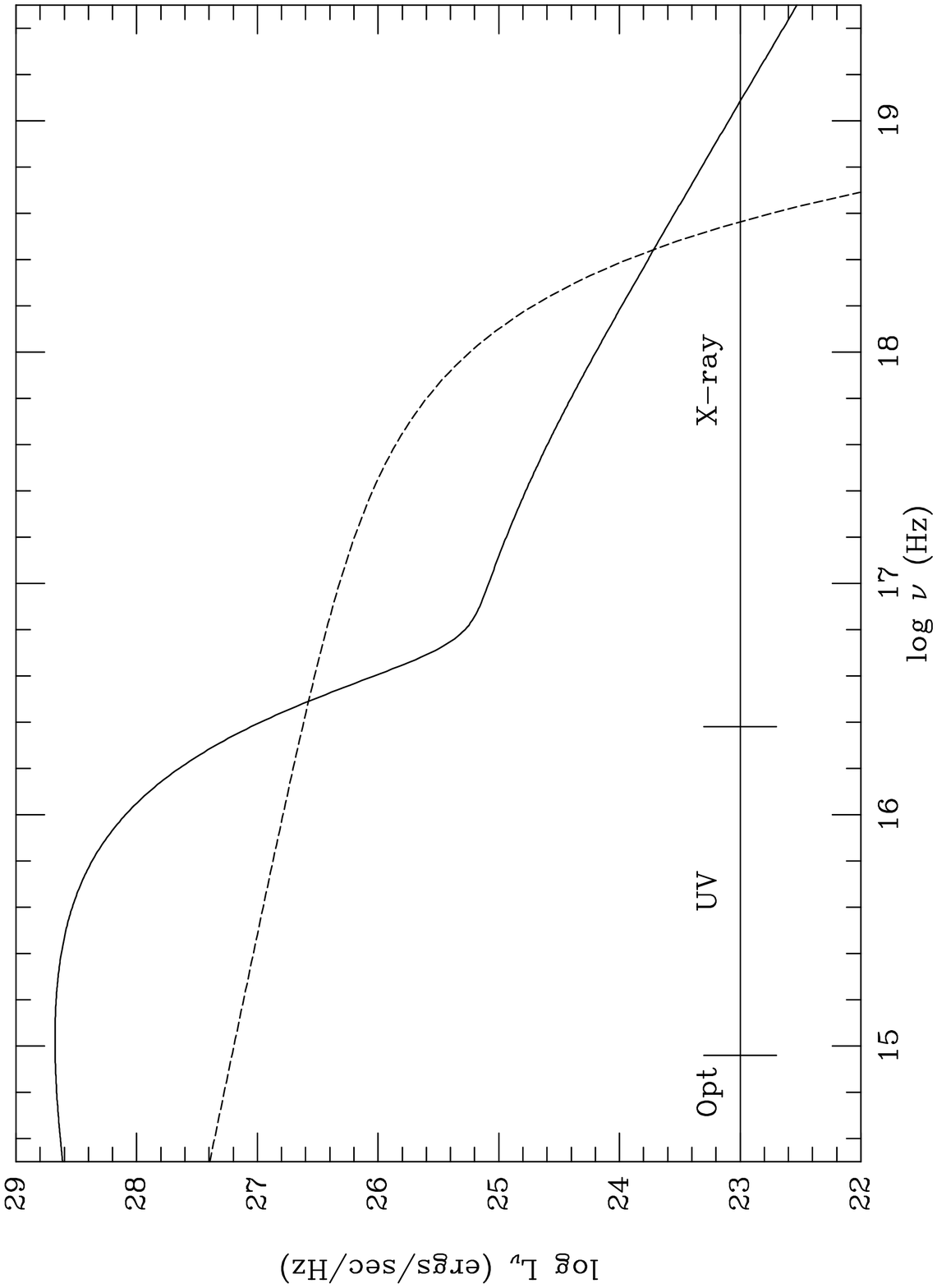}
\centerline{\null}
\centerline{\null}
\noindent\hskip2.76truein (a) \hskip2.94truein (b) \hfill
\vskip1.70in
\caption[]{(a) A typical model for the relativistic electron population
in a cluster of galaxies.
The lower energy electrons are due to all of the mergers in the cluster
history, while the high energy electrons are due to a small current
merger.
(b)
The IC spectrum from the same model (solid curve).
The dashed curve is a 7 keV thermal bremsstrahlung spectrum.\hfill
\label{fig:espect}
}
\end{figure}

While it is possible that the EUV emission may be thermal in
origin,\cite{F97,BLM}
I believe that it is more likely that this emission is due to
inverse Compton scattering (IC) of CMB photons by low energy
relativistic electrons.\cite{H97,BB,EB,SL}
In this model, the EUV would be produced by electrons with
energies of $\sim$150 MeV ($\gamma \sim 300$; Fig.~\ref{fig:espect}).
As noted above, these electrons have lifetimes which are comparable to
the Hubble time, and should be present in essentially all clusters.
This can explain why EUV emission is common.
To produce the EUV luminosities observed, one needs a population of
such electrons with a total energy of $\sim$$10^{62}$ ergs, which is
about 3\% of the typical thermal energy content of clusters.
This is a reasonable acceleration efficiency for these particles, given
that both the thermal energy in the intracluster gas and the relativistic
particles result from merger shocks.
The steep spectrum in going from EUV to X-ray bands is predicted by
this model
[Fig.~\ref{fig:espect}(b)],
it results from the rapid increase in losses ($\propto \gamma^2$)
for particles as the energy increases above $\gamma \sim 300$
(Fig.~\ref{fig:lifetimes}).

\begin{figure}[t]
\vskip2.30in
\includegraphics{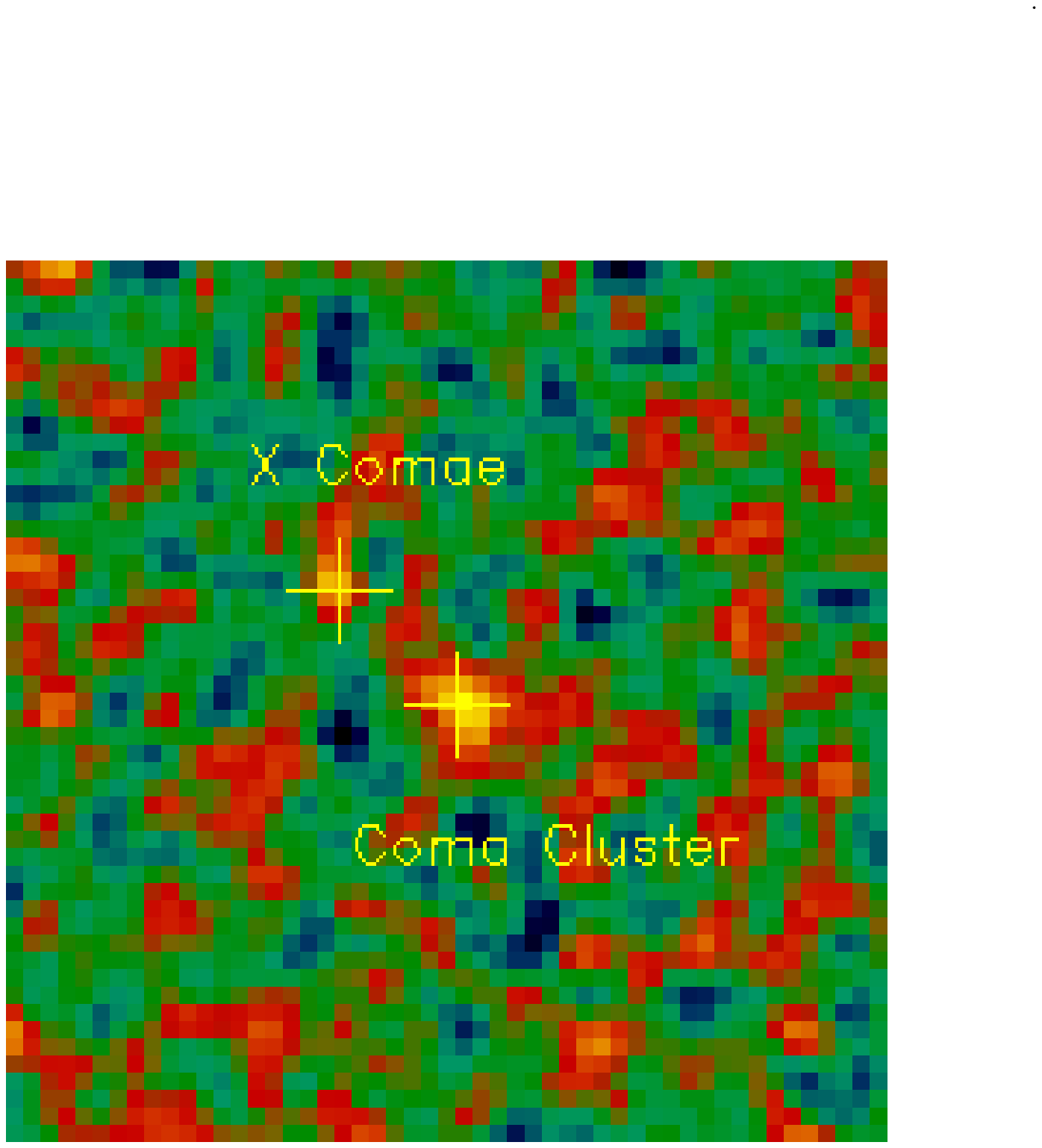}
\caption[]{A simulated INTEGRAL--IBIS image of the hard X-ray emission
from the Coma cluster in the 40--120 keV band.\cite{goldoni}
Such observations can separate the cluster emission from that of the nearby
AGN X-Comae, and can determine a crude radial distribution for the
emission.\hfill
\label{fig:integral}}
\end{figure}

The broad spatial distribution of the EUV is also naturally explained by the
density dependence of IC emission.
The thermal emission which produces the bulk of the X-ray emission in
clusters is due to collisions between thermal electrons and ions;
thus, it declines with the square of the density as the radius increases.
On the other hand, IC emission is due to collisions between cosmic ray
electrons and CMB photons;
since the CMB energy density is extremely uniform, IC EUV emission
varies with a single power of density, rather than density squared.
This simple difference can explain why the EUV is more extended than
the thermal X-ray emission.

\subsection{Hard X-ray Tails } \label{sec:nonthermal_hxr}

If clusters contain higher energy relativistic electrons with
$\gamma \sim 10^4$, these particles will produce hard X-ray (HXR) emission
by IC scattering, and radio synchrotron emission depending on the
intracluster magnetic field.
Since these higher energy electrons have short lifetimes, they should
only be present in clusters with evidence for a recent or ongoing merger.

Because of the short lifetimes of the electrons producing HXR IC emission,
the population of these particles should be close to steady-state.
The emitting electrons would have a nearly power-law energy distribution with
an exponent which is one power steeper than that of the accelerated
particles.\cite{ginzberg}
The HXR luminosity should be proportional to the energy input from
particle acceleration due to the cluster merger.
To a good approximation, the present day value of $L_{\rm HXR}$
(20--100 keV) is simply given by\cite{S1}
\begin{equation} \label{eq:hxr_lum}
L_{\rm HXR} \approx 0.17 {\dot{E}}_{\rm CR,e} ( \gamma > 5000 ) \, .
\end{equation}
where ${\dot{E}}_{\rm CR,e} ( \gamma > 5000 )$ is the total present rate of
injection of energy in cosmic ray electrons with $\gamma > 5000$.
If the relativistic electrons are accelerated in merger shocks, then
${\dot{E}}_{\rm CR,e} ( \gamma > 5000 ) $ should just be proportional to
the rate of energy dissipation in shocks.

Hard X-ray emission in excess of the thermal emission and detected as
a nonthermal tail at energies $\ga$20 keV has been seen in at least
two clusters.
The Coma cluster, which is undergoing at least one merger and which
has a radio halo, was detected with both BeppoSAX and
RXTE.\cite{FFea1,RGB}
BeppoSAX also has detected Abell~2256,\cite{FFea2}
another merger cluster with strong diffuse radio emission.
BeppoSAX may have detected Abell~2199,\cite{Kea}
although I believe the evidence is less compelling for this case.
A nonthermal IC hard X-ray detection of Abell~2199 would be surprising,
as this cluster is very relaxed and has no radio halo or relic.\cite{KS1}

The ratio of hard X-ray IC emission to radio synchrotron emission allows
one to determine the magnetic field in clusters.\cite{rephaeli79,FFea1,FFea2}
The present HXR detections lead to average magnetic fields which are
rather low ($\sim$0.2 $\mu$G)\cite{FFea1,FFea2})
compared to the values from Faraday rotation measurements\cite{Fea,CKB}

An alternative explanation of the hard X-ray tails is that they might be
due to nonthermal bremsstrahlung,\cite{ELB,Blasi00,dogiel,SK}
which is bremsstrahlung from nonthermal electrons with energies of
10--1000 keV which are being accelerated to higher energies.
However, it may be difficult to produce a sufficient nonthermal tail
on the electron distribution.\cite{petrosian}

The previous hard X-ray detections of clusters have been done with
instruments with very poor angular resolution.
Thus, they provide no information on the distribution of the hard X-ray
emission.
It would be very useful to determine if the hard X-ray emission is
localized to the radio emitting regions in clusters.
For clusters with radio relics, these might be associated with the
positions of merger shocks in the X-ray images.
Better angular resolution would also insure that the hard X-ray detections
of clusters are not contaminated by emission from other sources.
The IBIS instrument on INTEGRAL will provide a hard X-ray capability with
better angular resolution, and may allow the hard X-ray emission regions
to be imaged.\cite{goldoni}
Figure~\ref{fig:integral} shows a simulation of an IBIS observation
of the Coma cluster.

The predicted IC emission from nonthermal particles is much weaker than
the thermal emission in the central portion of the X-ray band
from about 0.3 keV to 20 keV
[Fig.~\ref{fig:espect}(b)],
However, if the IC emission is localized to merger shock regions,
its local surface brightness might be comparable to the thermal
X-ray emission.
A possible detection of localized IC emission associated with merger shocks
and radio relics has been claimed in Abell~85.\cite{bagchi}
It is possible that Chandra and XMM/Newton will find IC emission
associated with other merger shocks and radio relics.


\subsection{Radio Halos and Relics } \label{sec:nonthermal_radio}

A number of clusters of galaxies are known to contain large-scale
diffuse radio sources which have no obvious connection to individual
galaxies in the cluster.\cite{Gea}
They also have very steeply declining radio spectra.
These sources are referred to as radio halos when they appear projected
on the center of the cluster, and are called relics when they are found
on the cluster periphery (although they have other distinctive
properties).
The best known and studied radio halo is in the Coma cluster.\cite{Dea}
These radio halos and relics are relatively uncommon; about 40 are known
at the present time.
Although radio halos and relics have been studied for some time, lately the
number of known cases has increased dramatically.
This is largely due to the radio surveys (like NVSS and WENSS) and
new low frequency radio instruments.
Perhaps the largest number of new sources have come from the NVSS
survey and VLA pointed observation follow up effort by Giovannini,
Feretti, and collaborators.\cite{Gea}
Other new detections have involved other VLA surveys,\cite{OMV}
the VLA at 74 MHz,\cite{KPE}
SUMSS and MOST,\cite{Rea}
the ATCA,\cite{LHBA} and
WENSS.\cite{KS2}

\begin{figure}[t]
\vskip2.25in
\includegraphics{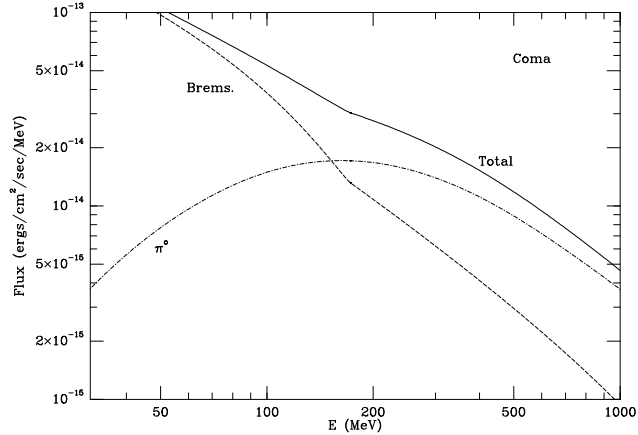}
\caption[]{The predicted gamma-ray spectrum for the Coma cluster, including
electron bremsstrahlung and $\pi^o$ decay from ions.\cite{S2}\hfill
\label{fig:gamma}}
\end{figure}

In all cases of which I am aware, they have been found in clusters which
show significant evidence for an ungoing
merger.\cite{Gea,Fer1,Fer2,SB,buote}
This suggests that the relativistic electrons are accelerated by merger
shocks.
In the early stages of mergers, halos are often found on the border
between the subclusters, where the cluster gas is first being shocked
(e.g., Abell 85).\cite{SR}
In more advanced mergers, more conventional centrally located halos
(e.g., Coma) and peripheral relics (e.g., Abell 3667) are found.
Abell~3667 provides a spectacular case of two bow-shaped radio relics
at opposite sides of a merging cluster,\cite{Rea}
and located at the positions where merger shocks are predicted.\cite{RBS}
Recent Chandra X-ray images of two clusters with radio halos and mergers
(Abell 665 and Abell 2163) show evidence for a simple connection between
the shocked regions and the radio emission.\cite{MV}

\begin{figure}[t]
\vskip3.47truein
\includegraphics{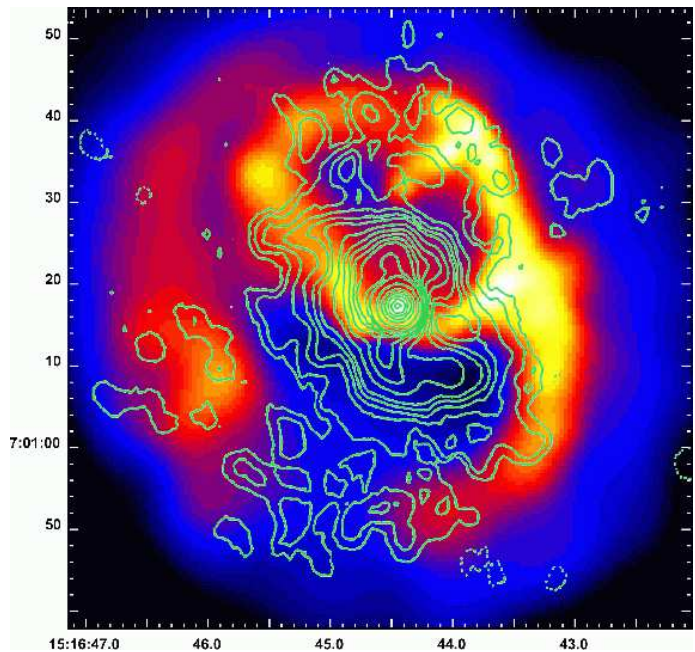}
\caption[]{The color image shows the central
$\sim$$80\arcsec$$\times$$80\arcsec$
region of an adaptively smoothed version of
the 0.3--10 keV Chandra image of the center of Abell~2052.\cite{blanton}
The green contours are the 6 cm radio image.\cite{burns}\hfill
\label{fig:a2052_radio}}
\end{figure}

\subsection{Predicted Gamma-Ray Emission} \label{sec:nonthermal_gamma}

Relativistic electrons and ions in clusters are also expected to produce
strong gamma-ray emission.\cite{EBKW,CB,BC,Bla,S2,DE}
The region near 100 MeV is particularly
interesting, as this region includes bremsstrahlung from the most common
electrons with $\gamma \sim 300$, and gamma-rays from ions produced by
$\pi^o$ decay.
Both of these processes involve collisions between relativistic particles
(electrons for bremsstrahlung, ions for $\pi^o$ emission)
and thermal particles, so they should both vary in the same way with
density in the cluster.
Thus, the ratio of these two spectrally distinguishable emission processes
should tell us the ratio of cosmic ray ions to electrons in
clusters.\cite{Bla,S2}

\begin{figure}[t]
\vskip3.45truein
\includegraphics{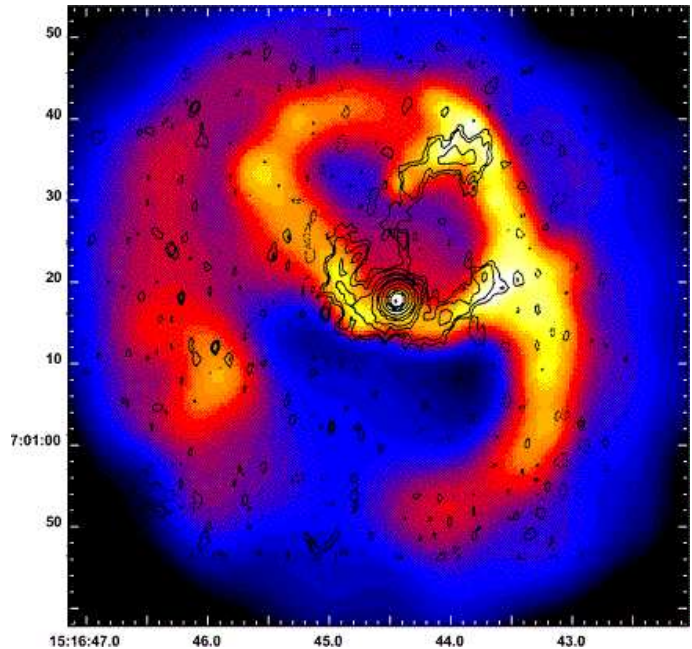}
\caption[]{Overlay of the H$\alpha$ + [N II] contours\cite{baum}
onto the central $\sim$$80\arcsec$$\times$$80\arcsec$
X-ray image of Abell~2052.\cite{blanton}\hfill
\label{fig:a2052_halpha}}
\end{figure}

Figure~\ref{fig:gamma} shows the predicted gamma-ray spectrum for the
Coma cluster, based on a model which reproduces the observed EUV, hard
X-ray, and radio emission.\cite{S2}
The observed upper limit from CGO/EGRET is $<$$4 \times 10^{-8}$
cts/cm$^2$/s for $E > 100$ MeV,\cite{Sea}
while the predicted value for this model is
$\sim$$2 \times 10^{-8}$ cts/cm$^2$/s.
The EGRET upper limit already shows that the ratio of ions to electrons
cannot be too large ($\la$30).\cite{Bla,S2}
The predicted fluxes are such that many nearby clusters should be easily
detectable with GLAST.

\section{Radio Source / Cooling Flow Interaction in Abell~2052}
\label{sec:a2052}

Although it is not really covered by my assigned topic,
I cannot resist the opportunity to show some new results from
our Chandra observation of the cooling flow cluster Abell~2052.\cite{blanton}
(Since this involves the interaction of the ``nonthermal'' radio source
with the X-ray gas, perhaps it just fits under my title.)
Abell~2052 is a moderately rich, cooling flow cluster at a redshift of
$z=0.0348$.
The central cD in this cluster hosts the powerful, complex radio source 3C~317.
Abell 2052 was observed with {\it Chandra} on 2000 September 3 for a total
of 36,754 seconds.
The smoothed X-ray image and radio contours are shown in
Figure~\ref{fig:a2052_radio}.
There is an X-ray point source coincident with the radio AGN.
The AGN lies just above an East-West bar of X-ray emission.
The extended radio emission corresponds with ``holes'' in the
X-ray emission, and the radio source is surrounded by a brightened
``shell'' of X-ray emission.
In the northern X-ray hole / radio lobe, there is a ``spur'' of
X-ray emission sticking into the hole.
The X-ray shell around the southern radio lobe is incomplete at the
bottom.
The X-ray hole / radio lobe structure is very similar to that seen with
Chandra in
Hydra A,\cite{Mc1}
Perseus,\cite{Fabian2000}
Abell~2597,\cite{Mc2}
and possibly RBS797.\cite{SCDSW}

The pressure in the X-ray-bright gas is nearly continuous with the
pressure of the surrounding gas.
There is no clear evidence for strong shocks.
Thus, it seems likely that the radio lobes are displacing and compressing
the X-ray gas, but are, at the same time, confined by the X-ray gas.
The minimum nonthermal pressure in the radio lobes is about an order of
magnitude smaller than that in the X-ray shell and surrounding thermal
gas.
Thus, it is likely that the usual equipartition arguments fail for this
(and presumably other) radio sources.
The radio lobes might be supported by pressure from
low energy relativistic electrons,
relativistic ions,
magnetic fields,
or very hot, low density thermal gas.

Contours of H$\alpha$ + [N II] emission\cite{baum}
are plotted over the smoothed X-ray
emission in Figure~\ref{fig:a2052_halpha}.
Optical emission lines are seen from the AGN, from the East-West bar, and
from the brightest portions of the shell, including the spur.
The compression of the X-ray shell by the radio source appear to have
promoted cooling, including to low temperatures
$\sim$$10^4$ K.
A deeper optical line image might reveal emission from the remainder of
the X-ray shell.

\section*{Acknowledgments}
I want to thank my collaborators
Liz Blanton,
Josh Kempner,
Maxim Markevitch,
Scott Randall,
Paul Ricker,
and
Alexey Vikhlinin
for all their help.
Liz Blanton, Paolo Goldoni, Josh Kempner, and Paul Ricker kindly provided
figures for this paper.
Support for this work was provided by the National Aeronautics and Space
Administration through Chandra Award Numbers
GO0-1119X, GO0-1173X, GO0-1158X, and GO1-2122X,
issued by the Chandra X-ray Observatory Center, which is operated by the
Smithsonian Astrophysical Observatory for and on behalf of NASA under contract
NAS8-39073.
Support also came from NASA XMM grants
NAG 5-10074 and NAG 5-10075.

\end{document}